\documentclass{PoS}

\def\bea{\arraycolsep .1em \begin{eqnarray}}
\def\eea{\end{eqnarray}}

\def\bea{\arraycolsep .1em \begin{eqnarray}}
\def\eea{\end{eqnarray}}
\def\no{\nonumber}

\def\s0#1#2{\mbox{\small{$ \frac{#1}{#2} $}}}
\def\0#1#2{\frac{#1}{#2}}

\title{Studies on X(4260) and X(4660) particles}

\ShortTitle{Studies on X(4260) and X(4660) particles}

\author{{Meng Shi} \\
        Department of Physics, Peking University, Beijing 100871, China\\
        E-mail: \email{shimeng1031@pku.edu.cn}}

\author{De-liang~Yao\\
        Department of Physics, Peking University, Beijing 100871, China\\
        E-mail: \email{yaodeliang@pku.edu.cn}}

\author{\speaker{Han-qing~Zheng}\\
        Department of Physics, Peking University, Beijing 100871, China\\
        E-mail: \email{zhenghq@pku.edu.cn}}

\abstract{Studies on the X(4260) and X(4660) resonant states in an
effective lagrangian approach are reviewed. Using a Breit--Wigner
propagator to describe their propagation, we find that the X(4260)
has a sizable coupling to the $\omega\chi_{c0}$ channel, while other
couplings are found to be negligible. Besides, it couples much
stronger to $\sigma$ than to $f_0(980)$:  $|g_{X\Psi
\sigma}^2/g^2_{X\Psi f_0(980)}|\sim O(10) \ .$  As an approximate
result for X(4660), we obtain that the ratio  of
$\frac{Br(X\rightarrow\Lambda_c^+\Lambda_c^-)}{Br(X\rightarrow\Psi(2s)\pi^+\pi^-)}\simeq
20$. Finally, taking X(3872) as an example, we also point out a
possible way to extend the previous
 method to a more general one in the effective lagrangian approach.}

\FullConference{Xth Quark Confinement and the Hadron Spectrum,\\
        October 8-12, 2012\\
        TUM Campus Garching, Munich, Germany}

\begin{document}
Many near threshold resonances, namely the X, Y, Z states,  have been
discovered in recent years experimentally~\cite{PDG2012}, which has
generated great interests in theoretical studies. For example, the
X(3872) state is very close to $D^0D^{0*}$ threshold, the Y(4260)
state locates very close to the $\omega\chi_{c0}$ and $DD_1(2420)$
thresholds, etc.. The understanding to these newly observed states
makes a severe challenge to the study of the heavy quark spectrum.

In this talk we will review our recent  studies of the Y(4260),
Y(4660) and X(3872) particles, using an effective lagrangian approach.
We will also point out a possible way to extend the previous
 method to a more general one in the effective lagrangian approach.
 We hope the new proposal for future studies can be helpful in   understanding   the
 formation of hadronic molecule in the heavy quark system and in
 distinguishing between a molecular state and a heavy quarkonium
 state.
\section{On X(4260) state}
The X(4260) state was firstly discovered by  BABAR Collaboration in
 2005~\cite{Babary4260} in initial state radiation (ISR) process, with $J/\psi\pi^+\pi^-$ in the final state. The mass and width
were found to be
 $M=4259\pm8(stat.)^{+2}_{-6}(sys.)$~MeV and
 $\Gamma=88\pm23(stat.)^{+6}_{-4}(sys.)$~MeV, and the branching ratio was given by
 $\Gamma_{e^+e^-}\times Br(X\to\pi^+\pi^-
 J/\psi)=5.5\pm1.0^{+0.8}_{-0.7}$~eV.
 This state has also been confirmed by CLEO~\cite{CLEOy4260a} and
 BELLE~\cite{Belley4260} experiments. On the other hand, it is puzzling that the X(4260) state is
 not found in the BES R-value
 measurement. Instead, there is only a dip structure in the energy
 region around 4.26GeV~\cite{bes}.
   In theory aspect many theoretical works have been
devoted to the study of the X(4260), and it is generally believed that the
existence of the X(4260) signals a degree of freedom beyond conventional
$\bar cc$ state. Many proposals have been made in the literature,
e.g., charmonium, $\chi_{c0}\rho^0$ molecule, $\omega\chi_{c1}$
  molecule,    $c\bar c g$ hybrid
state,  $\Lambda_c\bar\Lambda_c$ bayronium, $D_1\bar D$
  or $D_0\bar D^*_0$ molecule,
  etc.. There even exists the suggestion that the X(4260) may not even be a
  resonant state~\cite{nonresonant}. In the following we will
  however assume that X(4260) is a propagating Breit--Wigner state
  and the denominator of the X(4260) is parameterized  as,
\bea\label{X4260Den}
  D_X(q^2)=M_X^2-q^2-i{
  \sqrt{q^2}}(g_1k_1+g_2k_2+\Gamma(q^2)+{ \Gamma_0})\ ,
  \eea
where $M_X$ is the bare mass of the Breit-Wigner particle, $g_1$
($g_2$) denotes the coupling of X(4260) to the nearest channel
below (above) the pole position,  $k_1$ and $k_2$ are corresponding
channel momentum respectively. $\Gamma(q^2)$ denotes the partial
decay width to $J/\Psi\pi\pi$ and the constant width $\Gamma_0$
simulates other possible decay channels apart from those included
explicitly in Eq.~(\ref{X4260Den}). Most likely $\Gamma_0$ would
represent the (missing) open charm channels which are unobserved
experimentally.

\begin{figure}[h]
\begin{minipage}[l]{0.31\textwidth}
\includegraphics[width=1.0\textwidth]{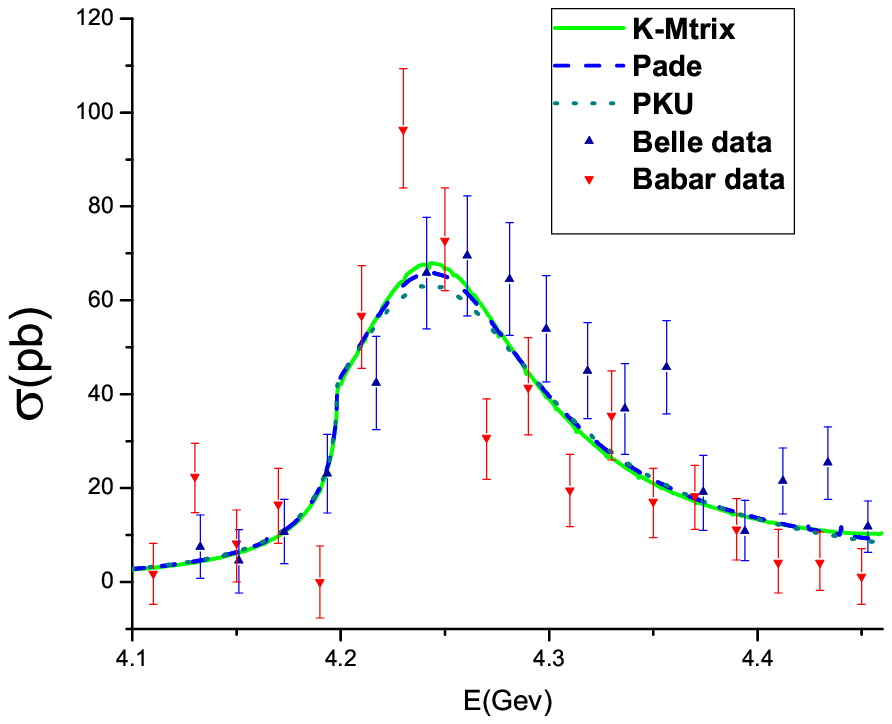}\label{fig1}
\end{minipage}
\begin{minipage}[r]{0.31\textwidth}
\includegraphics[width=1.0\textwidth]{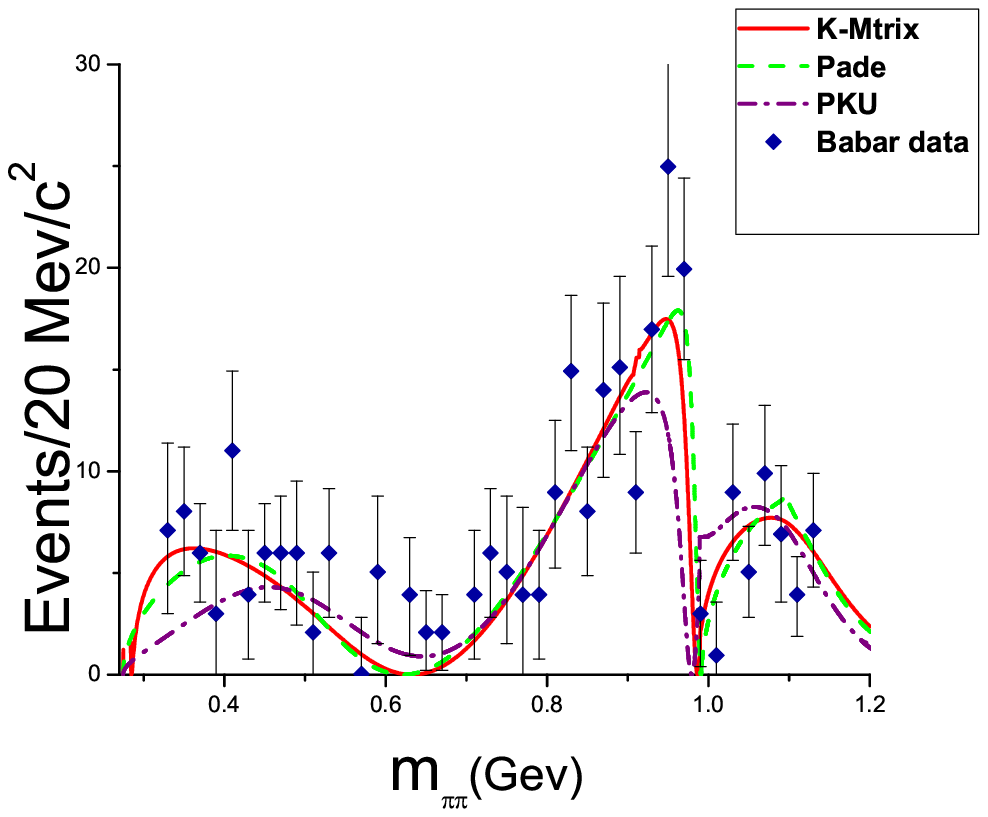}\label{fig2}
\end{minipage}
\begin{minipage}[r]{0.31\textwidth}
\includegraphics[width=1.0\textwidth]{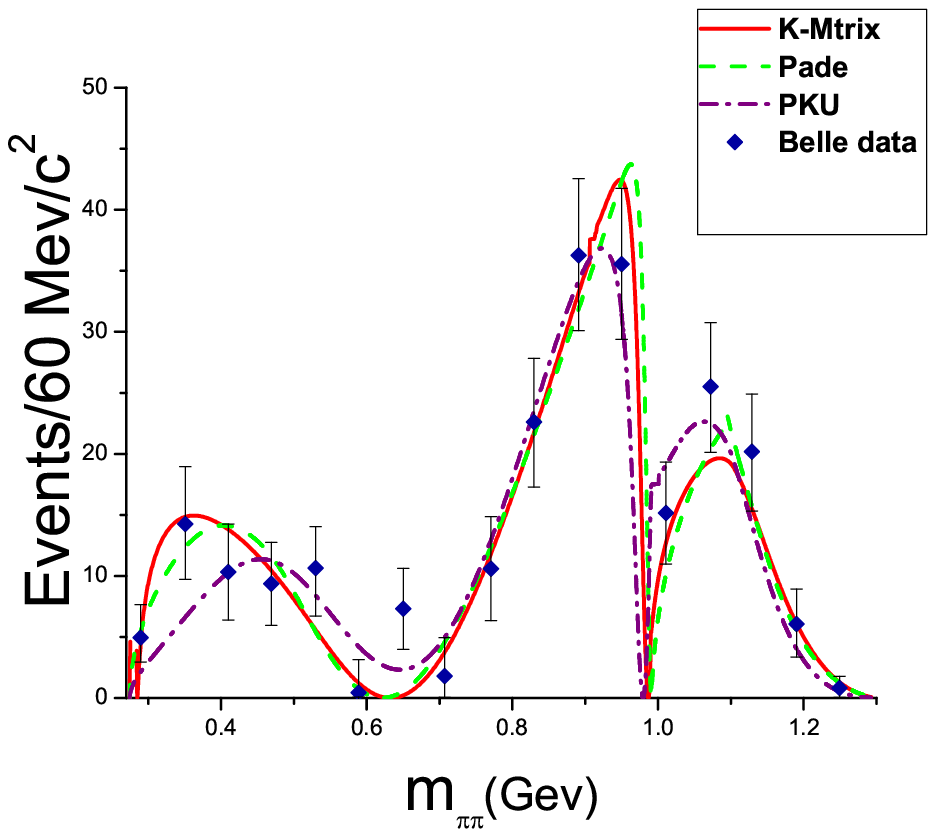}\label{fig3}
\end{minipage}
\caption{\label{4260fig} Left, the $J/\Psi\pi\pi$ cross section from
BABAR\cite{Babary4260} and BELLE\cite{Belley4260}; middle, the
$\pi\pi$ invariant mass spectrum from BABAR\cite{Babary4260}; right,
the $\pi\pi$ invariant mass spectrum from BELLE\cite{Belley4260}.}
\end{figure}

  We write down effective lagrangians describing its photoproduction and decay into $J/\Psi\pi\pi$ states:
\bea\label{eq1}
\mathcal{L}_{\gamma X}&=&g_0 X_{\mu\nu}F^{\mu\nu} \,\no\\
\mathcal{L}_{X\psi PP}&=&h_1 X_{\mu\nu}\psi^{\mu\nu} <
u_{\alpha}u^{\alpha} > +h_2 X_{\mu\nu}\psi^{\mu\nu}<\chi_+> +h_3
X_{\mu\alpha}\psi^{\mu\beta}<u_{\beta}u^{\alpha}> \ .
 \eea
The above lagrangian obeys chiral symmetry up to ${\cal O}(p^2)$
level and for more detailed explanation we refer to
Ref.~\cite{DSZ2012}. Eq.~(\ref{eq1}) is not enough yet to
appropriately describe the strong interactions of the I=0 $s$-wave
$\pi\pi$ final state. We therefore only use Eq.~(\ref{eq1}) to
calculate the tree level decay amplitude ${\cal A}^{tree}$, and
further improve the calculation  by making use of the couple channel
final state theorem~\cite{pennington}: \bea\label{MP} \mathcal {
A}_{1}&=&\mathcal {A}^{tree}_{1}\alpha_1(s) T_{11}(s)+\mathcal
{A}_{2}^{tree}\alpha_2(s) T_{21}(s)\
 ,\no\\
 \mathcal {A}_{2}&=&\mathcal {A}_1^{tree}\alpha_1(s) T_{12}(s)+\mathcal {A}_{2}^{tree}\alpha_2(s) T_{22}(s)\
 ,\no
 \eea
where the subscripts 1, 2 denote the $\pi\pi$ and $\bar KK$ channels,
respectively, and $\alpha_i$ are mild polynomials to be determined
by fit. For the $\pi\pi$, $\bar KK$ scattering $T$ matrix we chose
three solutions found in the
literature: Pad\'e~\cite{pade}, K-matrix~\cite{yumao}, PKU~\cite{pku2}.
The fit results are shown in Fig.~\ref{fig1} and
table~\ref{para4260V}. From numerical studies we draw the following
observations: A large coupling between X(4260) and $\omega\chi_{c0}$ is
obtained, while other couplings are found to be negligible; It is estimated that the X particle couples much stronger to
$\sigma$ than to $f_0(980)$:  $|g_{X\Psi \sigma}^2/g^2_{X\Psi
f_0(980)}|\sim O(10) \ .$
The value of $g_0$ given in table~\ref{para4260V} corresponds to $\Gamma(e^+e^-)=228$~eV. These numbers are in reasonable range
comparing with the BES bound given in Ref.~\cite{mo}.

One unresolved puzzle with respect to X(4260) is that, if it is a
$\bar cc$ state, it is hard to explain the absence of its decay into
open charm channels, e.g., $D\bar D$, $D^*\bar D$, etc.. Here in the
fit the constant width is found to be 50~MeV with sizable uncertainty. One possible explanation to this puzzle is that there exists
a  cancelation between contributions from $\gamma^*$ and $X$ to open
charms.
\begin{table}[hb]
\begin{center}
\begin{minipage}[c]{0.40\textwidth}
\begin{center}
 \begin{tabular}  {c c c c}
 \hline\hline
                           &   Fit I  (Pad\'e)       \\ \hline
$\chi^2_{d.o.f}$  &  $\frac{108.9}{93-14}$   \\
$g_0$(MeV)           &   9.984  $\pm$1.046           \\
$g_1$                &    { 0.608$\pm 0.094$}        \\

$M_X$(GeV)      &     4.263 $\pm$0.010          \\
{ $\Gamma_0$}(GeV)     &  0.051 $\pm$ 0.008      \\
 \hline\hline
 \end{tabular}
 \caption{\label{para4260V}Fit results assuming   X(4260) couples to
$\omega\chi_{c0}$;
  $\Gamma_0$  and background included in the fit.}
  \end{center}
\end{minipage}
\hspace{0.5cm}
\begin{minipage}[c]{0.45\textwidth}
\begin{center}
 \begin{tabular}  {c c c c}
 \hline\hline
                           &   Fit I  (Pad\'e)       \\ \hline
$\chi^2_{d.o.f}$  &  $\frac{{ 147.0}}{93-14}$   \\
$g_0$(MeV)           &   6.836  $\pm$0.245           \\
$g_1$                &    0.514$\pm 0.008$        \\

$M_X$(GeV)      &     4.2118 $\pm$0.012          \\
$\Gamma_0$(GeV)     &  0.017 $\pm$ 0.014      \\


\hline\hline
 \end{tabular}
 \caption{\label{para4260VI}Fit results assuming an { equal} X(4260) coupling to $\omega\chi_{c0}$ and $\bar D D_1$;
  $\Gamma_0$  and background included in the fit.}
   \end{center}
\end{minipage}
\end{center}
\end{table}
We find two poles located at $\sqrt{s}=4177.3-90.0i$~MeV on sheet III and
$\sqrt{s}=4.227.4-39.7i$~MeV on sheet IV respectively.

One possible explanation found in the literature is that the X(4260)
be a candidate of $c\bar c g$ hybrid state. In such a situation, the
X(4260) may couple strongly to $DD_1$ channel~\cite{hybrid} which is
however not supported by our fit. In table~\ref{para4260VI}, we list
the result by enforcing a equal coupling of $X\omega\chi_{c0}$ and
$X DD_1$. Comparing with the fit result in
table~\ref{para4260V} the total $\chi^2$ is increased considerably.


Through our numerical studies, we also found that, except for the
$X\omega\chi_{c0}$ coupling, there are  no signals for the X(4260)
coupling to other channels. Hence we exclude most of the molecular
assignment to X(4260). Assuming the occurrence of the cancelation
between $\gamma^*$ and X(4260) to open charm channels, we may
conclude from our numerical analysis that the X(4260) is mainly a
$c\bar c$ state renormalized by the $\omega \chi_{c0}$ continuum.
Our estimated $\Gamma_{e^+e^-}\simeq 228$eV, which is within the
upper limit set up from BES experiments~\cite{mo}. The
renormalization effect due to $\omega\chi_{c0}$ loop should be
important because a naive quark model calculation tends to give a
large value of  $\Gamma_{e^+e^-}$. A screening inter-quark potential
can lower the mass of $4^3S_1$ state down to 4273~MeV with a
$\Gamma_{e^+e^-}\simeq 970$~eV, S-D mixing may reduce this number by
half~\cite{LiBQ}, which is however still not small enough in
comparison with our estimate. Hence a sizable mixing with the
continuum is crucial in reducing the leptonic decay width. Notice
that the $\gamma^*$--X transition coupling $g_0$ obeys
 \bea
g_0^R=Z_X^{1/2}g_0^B\ ,
 \eea
  where $g_0^B$ denotes the value of $g_0$ at tree level -- value obtained from simple potential model
calculation without considering the continuum mixing, and  $g_0^R$
is the `renormalized' quantity  measured by experiments. The wave
function renormalization constant $Z_X$ is finite and calculable for
$s$-wave interaction in non-relativistic limit. To understand this
better let us consider a  simplified situation when X is a bound
state with respect to the $\omega\chi_{c0}$ channel then
\bea\label{epsilon} Z_X
=\frac{1}{1-\mathrm{Re}\,\Sigma'(\mu^2)}\simeq\frac{1}{1+\frac{g_1}{2\sqrt{2}}
\sqrt{\frac{m_R}{\epsilon}}}\ , \eea where we have let
$\mu=M_{th}-\epsilon$  and $\epsilon$ is the binding energy,
$m_R=\frac{M_\chi\,m_\omega}{M_\chi+m_\omega}=637$MeV. The
 loop correction leads to  a reduction of  the `tree level' value of $\Gamma_{e^+e^-}$ by a factor
$Z_X$.

\section{A brief comment on X(4660)}
\begin{table}[hb]
\begin{center}
 \begin{tabular}  {c c c c}
 \hline\hline
&   fit I  (Pad\'e)    &      fit II  (K-matrix)\\ \hline
$\chi^2_{d.o.f}$   &  $1.38$  & $1.00$    \\
$g_0$(MeV)     &  7.118 $\pm$0.633                  & 7.025  $\pm$  0.630           \\
$g_1$            &  2.155  $\pm$0.273           &  2.103   $\pm$0.275          \\
$M_X$(GeV)       &   4.659 $\pm$0.011          &   4.652 $\pm$0.010               \\
\hline\hline
 \end{tabular}
 \caption{\label{para4660}Parameters given by fit to X(4660) data.}
   \end{center}
\end{table}
The X(4660) state is observed to decay into $\Lambda_c\bar\Lambda_c$
and $\psi(2s)\pi\pi$.~\cite{4660} The effective lagrangian
describing the X(4660) interaction are the following:
 \bea
\mathcal{L}_{X\gamma}&=&g_0F_{X\mu\nu}F_{\gamma}^{\mu\nu},\\
\mathcal{L}_{\Psi'Xpp}&=&h_1F_{X\mu\nu}F_{\Psi'}^{\mu\nu}<u_{\rho}u^{\rho}>
+h_2F_{X\mu\nu}F_{\Psi'}^{\mu\nu}<\chi_+>
+h_3F_{X\mu\alpha}F_{\Psi'}^{\mu\beta}<u_{\beta}u^{\alpha}>\\
\mathcal{L}_{X\Lambda_c^+\Lambda_c^-}&=&g_1\bar\Psi_c\gamma^{\mu}\Psi_c
X_{\mu}\ .  \label{LagrangianLambdac}
 \eea
 The denominator of the Breit--Wigner X(4660) propagator is
 parameterized as
\bea
 D_X(q^2)=q^2-M_X^2+i\sqrt{q^2}(\frac{3}{2}\Gamma_{\psi(2S)\pi^+\pi^-}(q^2)+2\Gamma_{\psi(2S)K^+K^-}(q^2)+
  \Gamma_{\Lambda_c^+\Lambda_c^-}(q^2))\ .\label{propagator}
 \eea
Final state interactions
 among $\pi\pi$ and $\bar KK$ are also taken into account. The fit
 results are listed in table~\ref{para4660}. There are also two poles lying on
 the third and fourth Riemann sheet:
 $\sqrt{s}=4618.5-73.5i$~MeV (III) and $\sqrt{s}=4623.3-68.3i$~MeV (IV) for
 Pad\'e method; $\sqrt{s}=4616.2-69.1i$~MeV (III) and
 $\sqrt{s}=4624.0-60.7i$~MeV (IV) for K-matrix method.

 The ratio of
 $\frac{Br(X\rightarrow\Lambda_c^+\Lambda_c^-)}{Br(X\rightarrow\Psi(2s)\pi^+\pi^-)}$
are estimated as 23.9 for Pad\'{e} method and 19.3
 for K-Matrix method and the results are comparable to those in Refs.~\cite{Charmed_Baryonium, Y4660Meissner2}.
We point out  that the value of
 $g_0$ given in table~\ref{para4660} corresponds to
 $\Gamma_{e^+e^-}\simeq 102$~eV. Assuming the magnitude of \bea \Gamma_{e^+e^-}\times
 Br(X\rightarrow\Psi(2S)\pi^+\pi^-)\approx 5eV, \eea which will be similar with
 X(4260)~\cite{Babary4260}, the branch ratio of
 $Br(X\rightarrow\Psi(2S)\pi^+\pi^-)\approx 5\%$ can be derived. This means
 $Br(X\rightarrow\Lambda_c^+\Lambda_c^-)\approx 95\%$, in
 agreement with the ratio
 $\frac{Br(X\rightarrow\Lambda_c^+\Lambda_c^-)}{Br(X\rightarrow\Psi(2s)\pi^+\pi^-)}\approx 20$.

\section{Future improvement and outlook}
\begin{figure}[!ht]
\centering
\includegraphics[width=0.45\textwidth,height=4cm]{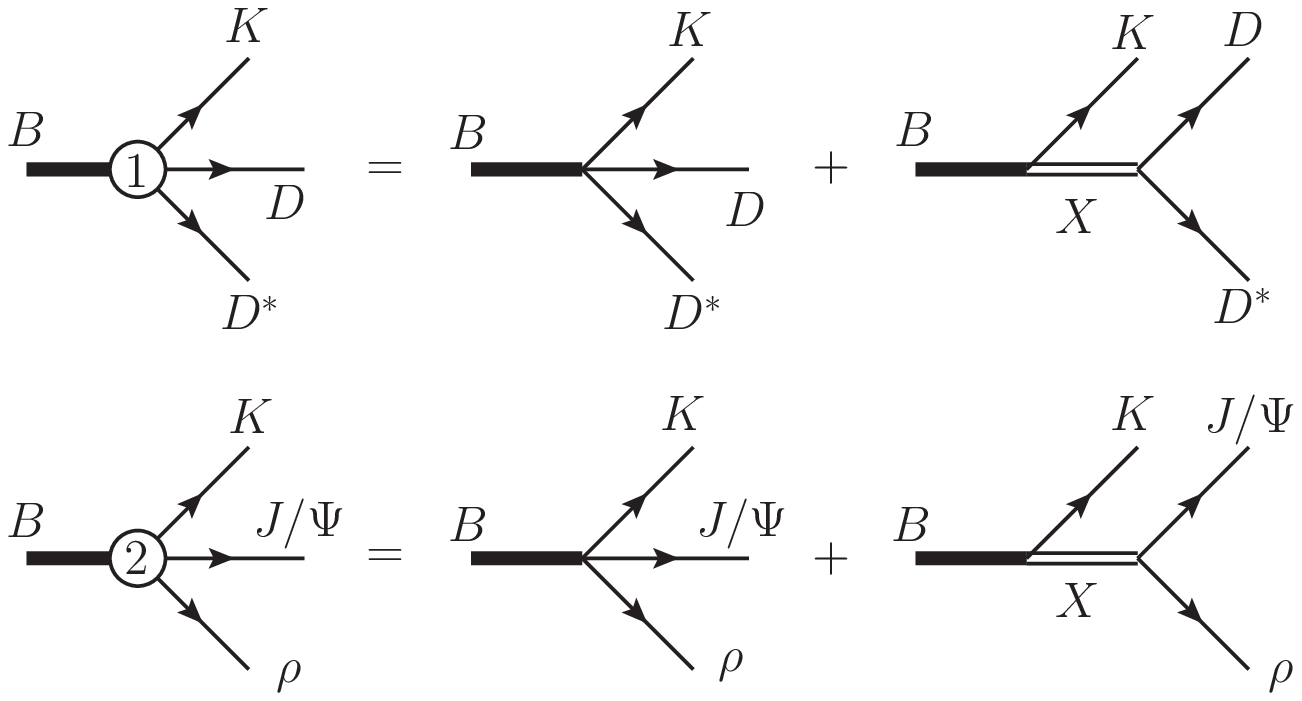}
\includegraphics[width=0.45\textwidth,height=4cm]{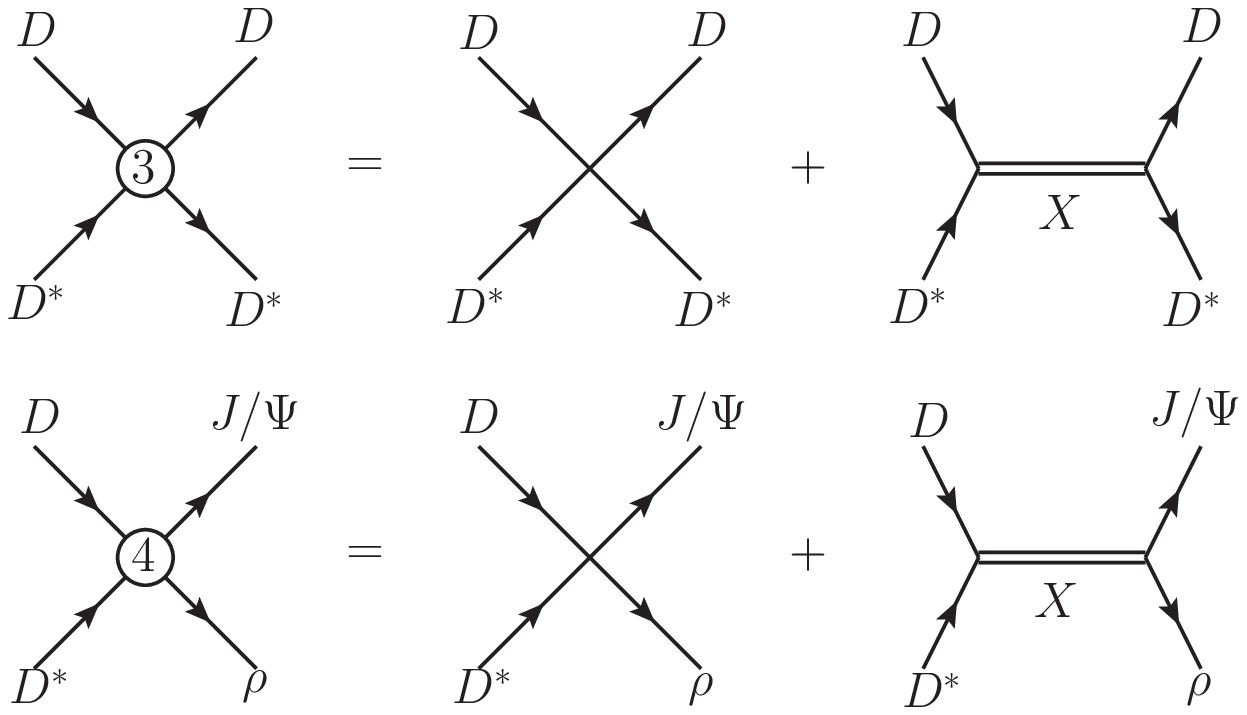}
\caption{Vertices appeared when studying X(3872).}\label{X3872vertex}
\end{figure}
\begin{figure}[!ht]
\centering
\includegraphics[width=0.70\textwidth]{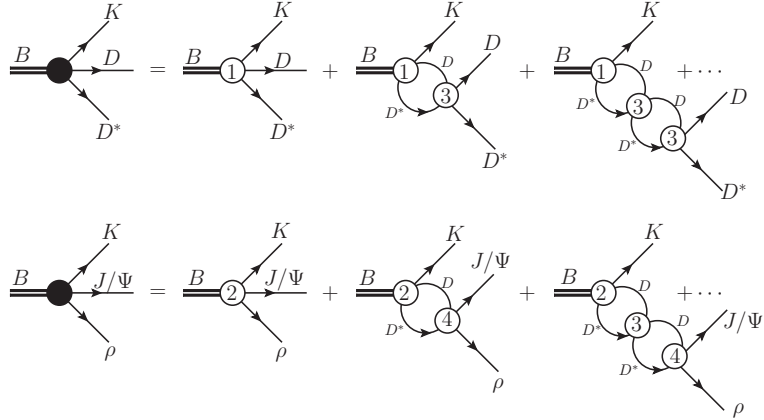}
\caption{The sum of bubble diagrams.}\label{X3872bubblediagram}
\end{figure}

In above we have performed a numerical analysis on the near threshold
resonances X(4260) and X(4660). However, one of the key assumption
we implicitly made in our analysis is that they propagate as a
particle, i.e., a Breit--Wigner  propagator is used to describe
their propagation. This assumption can in principle be examined and
tested by including more complicated dynamics. Taking the X(3872)
particle for example, as shown in figures \ref{X3872vertex}--\ref{X3872bubblediagram}, one can
include the final state interactions between $D^*\bar D$ and sum
up the bubble chains. The resummation of the diagrams shown in
figures~\ref{X3872vertex} and~\ref{X3872bubblediagram} is  an
approximation but becomes exact in the non-relativistic limit. In
this situation a molecular type state may be generated from the
bubble chain. Such a pole might interact with the bare Breit--Wigner
particle and the final physical picture might be then determined by the
competition of the two different type of poles. A study along this
direction is underway.

\noindent {\bf Acknowledgements:} This work is supported in part by
National Nature Science
Foundations of China under contract number  10925522 
 and
11021092.

\end{document}